\documentclass[12pt,preprint]{aastex}
\usepackage{emulateapj5}

\begin{document}

\title{Formation of globular clusters
with internal  abundance spreads
in r-process elements: strong evidence for
prolonged star formation}

\author{Kenji Bekki} 
\affil{
ICRAR,
M468,
The University of Western Australia
35 Stirling Highway, Crawley
Western Australia, 6009, Australia
}

\and

\author{Takuji Tsujimoto}
\affil{
National Astronomical Observatory of Japan, Mitaka-shi, Tokyo 181-8588, Japan}

\begin{abstract}
Several globular clusters (GCs) in the Galaxy
are observed to show internal abundance spreads
in $r$-process elements (e.g., Eu). 
We here propose a new scenario which explains the origin
of these GCs (e.g., M5 and M15).
In this scenario, stars with no/little abundance
variations first form from a massive molecular cloud (MC).
After all of the remaining gas of the MC
is expelled by numerous supernovae,
gas ejected from asymptotic giant branch stars 
can be accumulated in the central region of the GC to form a
high-density intra-cluster medium (ICM).
Merging of  neutron stars then occurs to eject 
$r$-process elements, which can be efficiently trapped in
and subsequently mixed with the ICM.
New stars formed
from the ICM can have  $r$-process abundances 
quite different from those of earlier generations of stars within the GC.
This scenario can explain both (i) why $r$-process elements 
can be trapped within GCs  and (ii) why GCs with internal  abundance spreads
in $r$-process elements do not show [Fe/H] spreads.
Our model shows that (i)
a large fraction of Eu-rich stars can be seen in Na-enhanced stellar populations
of GCs, as observed in M15,
and (ii)  why most of the Galactic GCs
do not exhibit such internal abundance spreads.
Our model demonstrates that the observed internal
spreads of $r$-process elements in GCs
provide strong evidence for prolonged star formation
($\sim 10^8$ yr). 
\end{abstract}

\keywords{
globular cluster: general --
galaxies: star clusters: general --
galaxies: stellar content --
stars:formation
}

\section{Introduction}

Recent photometric and spectroscopic investigation of
the Galactic globular clusters (GCs) has confirmed that the GCs exhibit
multiple stellar populations (see, e.g., Gratton et al. 2012 for a recent review).
Individual GCs show internal abundance spreads in different elements: light 
elements (e.g., Carretta et al. 2009; C09),  $s$-process (e.g., Marino et al. 2009),
C+N+O (e.g., Yong et al. 2015), $r$-process (e.g., 
Snedin et al. 1997; Roederer 2011, R11; Sobeck et al. 2011, S11;
Worley et al. 2013; W13),  helium (e.g., Piotto et al. 2005), and Fe (e.g.,
Freeman \& Rodgers 1975). 
Furthermore, the vast majority of the Galactic GCs investigated so far 
possess clear anti-correlations between light elements (e.g., CNO; C09).
Such internal abundance spreads have also been observed 
for GCs in 
the Large and Small Magellanic Clouds (e.g., Mucciarelli et al. 2009;
Niederhofer et al. 2016),
and the Galactic dwarf satellites (e.g.,
Larsen et al. 2014).
A key question related to GC formation is why GCs show both (i)  ubiquitous
anti-correlations between light elements (e.g., the Na-O anti-correlation)
and (ii) a diversity in the degrees of chemical abundance inhomogeneity
(e.g., only some GCs with [Fe/H] spreads).

Previous theoretical  models of GC formation tried to reproduce
the apparently universal anti-correlations between
O and Na and between Mg and Al observed in the Galactic GCs
(e.g., Fenner et al. 2004; Bekki et a 2007;
D'Ercole et al. 2008, 2010;  Ventura et al. 2016).
D'Antona, \&  Caloi (2004) discussed the origin of $Y$ (helium abundance spread)
in NGC 2808 based on the self-enrichment by massive AGB stars.
The origin of GCs with abundance spreads in Fe and $s$-process
elements has been recently discussed in the context of 
GC merging in their host dwarf galaxies (Bekki \& Tsujimoto 2016).
However,  
the origin of internal abundance spreads in $r$-process elements
have not be discussed extensively  so far,
though not only M15 and M92 (e.g., Snedin et al. 1997;
Roederer \& Snedin 2011) but also several others
(e.g., M5 and NGC 3201) are now observed to have
such abundance spreads (e.g., R11).
Tsujimoto \& Shigeyama (2014, TS14) suggested that 
efficient accretion 
of $r$-process elements ejected from merging between neutron stars
(neutron star merging; NSM)  onto some fraction of 
(i.e., not all of)
the stars in a GC can introduce a large star-to-star variation in
the abundances of $r$-process elements within the GC.

If NSM ejecta can be mixed with an intra-cluster medium (ICM) 
and finally used for secondary star formation in a GC,
then large star-to-star abundance spreads 
in $r$-process elements (e.g., [Eu/H]) can be expected.
A key question here is whether the NSM ejecta with very high ejection speed
($10-30$\% of the speed of light) can be stopped by the ICM
within the forming GC ($R<10$ pc).
Recently Komiya \& Shigeyama (2016) have shown that ejecta from NSMs
can be stopped through interaction with the interstellar medium (ISM),
because the ejecta can lose kinetic energy and momentum through 
interaction with the ISM. 
Recent numerical simulations have shown that a large amount
of AGB ejecta can be accumulated in the central regions 
of forming GCs to form very high density gaseous regions 
with $\rho_{\rm g}=10^5$ cm$^{-3}$ (Bekki 2017a, b, B17a,b).
If a single NSM occurs in such a high-density gaseous region,
then the ejecta is highly likely to be trapped and mixed with
the AGB ejecta.
New stars formed from such mixed gas can have chemical abundances of
$r$-process elements that are quite different from those of stars formed
in initial starburst at GC formation.
Thus, it is possible that secondary
star formation from AGB ejecta mixed with gas from a single NSM
can explain the large internal  abundance variations
observed in several Galactic GCs (e.g., M5 and M15).
This `NSM' scenario has not been explored by previous theoretical works.

The purpose of this paper is to investigate the origin
of GCs with internal abundance spreads in $r$-process elements
based on the NSM  scenario.
We particularly investigate the following questions: (1) under what physical conditions NSM ejecta can be trapped
within  forming GCs,  (2) what fraction of NSM ejecta can be trapped
within GCs and used for secondary star formation, and 
(3) whether the degrees of internal abundance spreads depend on the physical
properties of GCs (e.g., initial total masses). 
To do so, we use both analytical models and numerical simulations
of GC formation from star-forming molecular clouds. 
We provide predictions of (i) the number of GCs with such internal abundance spreads
(e.g., $\Delta {\rm [Eu/H]}$), (ii)  the bimodal distributions of [Eu/H] in GCs,
and (iii) possible correlations between [Na/Fe] and [Eu/H].

The plan of the paper is as follows.
We discuss the possibility of NSM ejecta being trapped
by the ICM of forming GCs using analytical models 
in \S 2.
We derive the possible internal spreads of [Eu/H] among GC stars
by assuming secondary star formation from NSM ejecta mixed with the ICM
in \S 3.
We present the results of numerical simulations of GC formation 
to discuss whether the ICM of forming GCs can be as high as the
required density of the ICM for trapping the NSM ejecta in \S 4.
Based on these results,
we provide several predictions of the NSM scenario
in \S 5.
We summarize our  conclusions in \S 6.
In order to discuss the present results, we show some observational results
(W13)
in Appendix A: the distribution of [Eu/H] and a correlation between
[Eu/H] and [Na/Fe] for M15.

\section{The scenario}

\subsection{Ruling out supernovae as a source of $r$-process elements in GCs}

There are two possible sites for the production
of  $r$-process elements, i.e., SNe and NSMs. 
Therefore,
the following two possible scenarios for the origin of the internal chemical
abundance spreads of $r$-process elements in GCs are promising. One is that
the intra-cluster medium (ICM)
of a GC-forming molecular cloud (MC)  is chemically polluted by ejecta
from several supernovae (SNe) so that new stars formed from the polluted gas
can have [Eu/H] different from those of stars formed from original gas
(`SN scenario').
The other is that new stars can be formed from  gas ejected from NSMs much later
than the initial burst of star formation in GCs (`NSM scenario'). 
In the SN scenario,  all stars with different [Eu/H] should be formed
before gas is completely expelled by SNe (i.e., $< 10^7$ yr),
which means that GCs are simply a single generation of stars.
Since the delay time ($t_{\rm delay}$)
distributions of NSMs has an extended distribution 
($10^7 \le t_{\rm delay} \le 10^{10}$ yr)
with a peak around $t_{\rm delay}=3 \times 10^7$ yr
(e.g., Dominik et al. 2012),
secondary star formation in the NSM scenario can  occur
after all gas is removed from the GC-forming MC. This means that prolonged star
formation is required in the NSM scenario.

We can rule out the SN scenario as follows.
If SNe are the site of heavy r-process elements such as Ba and Eu, 
they inevitably produce {\rm both} light (Y, Sr, etc) and heavy (Ba, Eu,
etc) r-process elements.
In M15, the stars which are enhanced in Ba and Eu abundances
do not exhibit any enhancement in Sr.
In other words, a large scatter is seen only
in Ba and Eu while no spread in light r-process elements is found
(e.g., S11).
This observational result strongly  suggests
that SNe are not associated with the observed internal spreads in $r$-process
elements of GCs (M15).
Furthermore, the observed lack of metallicity spreads in GCs 
([Fe/H]$<0.05$ dex; C09) is inconsistent with the SN
scenario, because if the ICM of forming GCs is polluted by $r$-process
elements, then other elements (e.g, Mg and Fe) are also polluted
to a large extent, ending up with
rather large metallicity spreads.
We thus rule out this SN scenario and accordingly discuss exclusively
the  NSM scenario in the present study.

\subsection{Secondary star formation in mixed gas from AGB stars and NSMs}

Figure 1 describes the new NSM scenario of GCs with
internal abundance spreads in $r$-process elements (e.g., [Eu/H]).
In the new scenario, the FG (first generation) stars  are formed from
a GC-forming molecular cloud. After all massive
stars  with the  masses ($m$) larger
than $9 M_{\odot}$ exploded as SNe, gas ejected from AGB stars
begins to be accumulated into the central region of the FG stellar system.
The formation of SG (second generation) stars 
in the central high-density gaseous region is possible due to accretion
of AGB ejecta.
One NSM occurs during this accretion phase when
the density of intra-cluster medium (ICM) in the central FG system
becomes quite high ($\rho_{\rm icm} > 10^4 M_{\odot}$).
As a result of this, the ejecta from the NSM
can be trapped by the ICM very efficiently, owing to interaction
between the NSM ejecta and hydrogen atoms. New stars 
are then formed from the ICM mixed with the NSM ejecta,
so that [Eu/H] of the new stars can be much higher than those of
FG stars. It should be stressed here that 
SG stars formed before the NSM can have almost identical [Eu/H] as
FG stars.

If the NSM occurs after the AGB ejecta are removed from a GC by some physical
processes, such as expulsion of the gas by SNIa  and ram pressure
stripping of the gas by the Galactic hot halo gas or by the warm gas
of the GC-host dwarf galaxy, then such a GC cannot show internal
spreads in $r$-process elements. This is because there is no gas
that can stop the $r$-process elements ejected from an NSM in the forming
GC. Accordingly, the timing of an NSM within a forming GC
is quite important as to whether the GC can finally have abundance spreads
in $r$-process elements. 
Since neutron stars are the outcome of deaths of massive
stars ($>10 M_{\odot}$ for which stars finally become SNe), 
NSMs  occur only after SNe in forming GCs. 
Therefore, 
gas ejected
from SNe should be all removed before the formation of SG stars.
Otherwise, this scenario cannot explain the observed lack of
abundance spreads in [Fe/H] among GCs with internal spreads
in $r$-process elements. 
Given the wide range of delay-time distributions for NSMs 
(e.g., Dominik et al. 2012),
it is possible that  NSM events can occur more than 1 Gyr after the 
formation of FG stars in GCs. This could explain why only a fraction
of old GCs shows significant abundance spreads in $r$-process elements
(e.g., R11, W13).

In this scenario, FG and SG stars formed before NSMs should 
show smaller [Eu/H] than SG stars formed after NSMs. As discussed
later in this paper, high [Eu/H] can be seen in SG stars with higher
[Na/Fe] in M15, which is consistent with the new scenario.
SG stars with very high [Na/Fe] formed from gas ejected by  massive AGB stars
($m\sim [8-9] M_{\odot}$)  
might be less likely to have high [Eu/H],
because the time interval between the onset of AGB phase
for such intermediate-mass ($m\sim [8-9] M_{\odot}$)  stars
and the last (lowest mass) SNe is very short:
fine-tuning is required for the epoch of an NSM.
The wide spread of [Eu/H] in SG stars with different [Na/Fe]
is expected in this scenario, if one NSM occurs when intermediate-mass
stars ($[3-9] M_{\odot}$) enter into the AGB evolutionary stage.

A key question in this scenario is how much gas (ICM) is required to
stop $r$-process elements ejected from NSMs, so that the NSM ejecta
can be trapped in the central regions of GCs.
Komiya \& Shigeyama (2016) analytically investigated 
how $r$-process elements from an NSM can lose their initial kinetic
energy through interaction with neutral hydrogen.  They estimated how
long the $r$-process elements can travel before they lose all
of their kinetic energy through Coulomb scattering, and found that
the `stopping length' ($l_{\rm s}$) is described as follows:
\begin{equation}
l_{\rm s}=2.6 (\frac{ n_{\rm HI} }{ 1 {\rm cm^{-3}} })^{-1} {\rm kpc},
\end{equation}
where $n_{\rm HI}$ is the number density of hydrogen atoms.
This $l_{\rm s}$ should be as small as the core sizes of forming GCs
($\sim [2-3]$ pc) where secondary star formation should occur.
It should be noted here that Tsujimoto et al. (2017) derived 
$l_{\rm s}$ using the observed abundances of $^{244}$Pu of pre-solar
grains, and pointed out that $l_{\rm s}$ can be significantly
smaller than that derived by Komiya \& Shigeyama (2016). Therefore,
$l_{\rm s}$ in the above equation is  an upper  limit for $l_{\rm s}$.

Figure 2 shows $l_{\rm s}$ as a function of the total mass of AGB ejecta
($M_{\rm gas}$)
for a given size of gas sphere ($R_{\rm gas}$). Here the AGB ejecta
is assumed to form a uniform gaseous sphere just for simplicity of
discussion. Clearly, if $M_{\rm gas} >10^4 M_{\odot}$ 
and $R_{\rm gas}<3$ pc, then $l_{\rm s}$ can be smaller than 3 pc.
This required $M_{\rm gas}$ is reasonable for GCs with
initial total masses ($M_{\rm gc,0}$) of $\sim 10^5 M_{\odot}$,
because $\sim 10$\% of the masses in AGB stars can be ejected through
stellar winds (e.g., Bekki 2011).
 If $M_{\rm gas} \sim 10^{5.5} M_{\odot}$,
then $r$-process elements can be trapped by the gas that is more
diffusely distributed (i.e., $R_{\rm gas} \sim 10$ pc).
These results imply that retaining NSM ejecta in the central regions
of forming
GCs are highly likely after gas ejected from AGB stars is accumulated
in the central regions.

If an NSM occurs within the gaseous sphere of a FG stellar system
with a mass of $M_{\rm FG} \ge 10^5 {\rm M}_{\odot}$ and $R_{\rm gas}<3$pc,
then almost all of the NSM ejecta can be retained by the gas.
If an NSM occurs outside the gaseous sphere of the FG stellar system ,
then only a fraction of the NSM ejecta can be trapped by the gas
(i.e., AGB ejecta). Figure 3 shows the fraction of NSM ejecta retained
by a GC ($f_{\rm ret}$),  as a function of the distance of the NSM from
the center of the GC ($r_{\rm gc}$),
for three $R_{\rm gas}=1$, 3,
and 10pc.
Since the flux of mass ejected from an NSM  falls as $r_{\rm g}^{-2}$,
$f_{\rm ret}$ can be rather small outside $R_{\rm gas}$.
It is likely that only a fraction
of NSM ejecta can be retained in GCs,
because NSMs are likely to occur in the outer regions of GCs  ($R>3$ pc,
where more stars can exist)
than the central regions.

We can estimate (i) the [Eu/H] of stars formed from NSM ejecta
mixed with AGB ejecta in a GC and 
(ii) internal abundance spreads of [Eu/H] among FG and SG stars 
($\Delta$[Eu/H])
in the GC for a given $f_{\rm ret}$ by using the following  simple model for 
mixing of AGB and NSM ejecta.
We assume that the AGB and NSM ejecta can be mixed uniformly so that
the final abundance of Eu depends on the total mass of AGB ejecta 
for a given yield of Eu from NSM events.
The total mass of gas ejected from one NSM ($M_{\rm nsm}$) is $\sim 0.01 M_{\odot}$
and the mass fraction of Eu in the ejecta ($f_{\rm Eu,0}$) is $10^{-2}$ (TS14). 
Using these numbers, we can calculate the chemical abundance of Eu 
($f_{\rm Eu}$) as follows:
\begin{equation}
f_{\rm Eu}= \frac{ f_{\rm Eu, 0}M_{\rm nsm}+f_{\rm Eu, agb}M_{\rm gas} }
{ M_{\rm agb} },
\end{equation}
where $M_{\rm gas}$ is
the total mass of AGB ejecta 
and $f_{\rm Eu, 0}$ and $f_{\rm Eu, agb}$
are the mass fractions of Eu in the NSM and AGB ejecta, respectively.
We here ignore the total mass of NSM ejecta in the denominator, because it
is too small in comparison with $M_{\rm gas}$.
We can convert this $f_{\rm Eu}$ into the final [Eu/H]
of the mixed gas (${\rm [Eu/H]_f}$)  by assuming the solar abundance
of Eu ($3.8 \times 10^{-10}$) for a given initial [Eu/H] 
(${ \rm [Eu/H]_i }$) of AGB ejecta.
The [Eu/H] spread is thus estimated as follows:
\begin{equation}
\Delta{\rm [Eu/H]}= {\rm [Eu/H]_f - [Eu/H]_i} .
\end{equation}

Figure 4 shows 
$\Delta$[Eu/H] for a metal-poor GC with ${ \rm [Eu/H]_i } = -2$
as a function of $M_{\rm gas}$ for three different $f_{\rm ret}$.
Clearly $\Delta$[Eu/H] for this metal-poor GC
can be quite large ($\ge 1$ dex) even for
$f_{\rm ret}=0.01$, if $M_{\rm gas} \le 10^{4.4} M_{\odot}$.
The GC with $M_{\rm gas} \le 10^{5} M_{\odot}$ has 
$\Delta$[Eu/H] much larger than the observed one for M15 ($\sim -1$).
A particular combination of $M_{\rm gas}$ and $f_{\rm ret}$ is
required for the observed $\Delta$[Eu/H]
to be reproduced. 
Since  GCs with $M_{\rm gc,0}$ can have 
$M_{\rm gas} \sim  10^{4} M_{\odot}$,
these results suggest that only a small fraction of NSM ejecta
should be mixed with ICM and retained in the GCs.
If $M_{\rm gc, 0}$ is significantly larger than 
$2\times 10^5 M_{\odot}$ (typical present-day GC mass), then
$f_{\rm ret}$ can be larger.
These results combined with those in Figure 4 imply that GCs
with internal abundance spreads of $r$-process elements experience
only one NSM event
in the outer part of the  GCs in their early phases of formation.
A future numerical study will investigate
where NSM events can occur in forming GCs 
with AGB ejecta.

\section{Analytical models for [Eu/H] spreads}

Observational study of [Ba/H] distributions of 
GCs with internal [Ba/H] spreads showed a bimodal distribution
of [Ba/H] in M15 (W13), and the [Eu/H] distribution
in the data by W13 and S11
shows such a  bimodality in the [Eu/H] distribution for M15 (see Appendix A).
It is thus important
to discuss whether and how the observed bimodality can be achieved in
the present scenario of GC formation.
Guided by recent results from hydrodynamical simulations
of turbulent diffusion (Greif et al. 2009),
we consider that
$r$-process elements ejected from NSMs can be spread over the ICM of a forming
GC through diffusion processes within a short timescale ($<10^6$ yr).
We adopt a working hypothesis that the chemical abundances of Eu can be 
different in different regions of the ICM owing to turbulent diffusion.
Although  gaseous regions close to
an NSM event  can have very high [Eu/H] initially,  the Eu abundances can become
progressively lower as  time passes  owing to diffusion.
We accordingly consider that the following functional form (the Green function)
described in Greif et al. 2009)
for the  distribution of [Eu/H] for stars formed from NSM ejecta mixed
with AGB ejecta:
\begin{equation}
N( {\rm [Eu/H] } )= \frac{ N_0 }{ {(2\pi \sigma^2)}^{3/2} } 
\exp( \frac{ - {(Z-Z_{\rm m})}^2 }{ 2 \sigma^2 }),
\end{equation}
where $N_0$ is the normalization constant,
$\sigma$ is the dispersion of [Eu/H],
$Z$ represents [Eu/H] (just for convenience), and 
$Z_{\rm m }$ is the mean value of [Eu/H].

These means and dispersions in [Eu/H]
are different between the stars formed from original gas of a GC-forming
MC (FG) 
and those from NSM ejecta mixed with AGB ejecta (SG).
If SG stars are formed from AGB ejecta before NSMs occur,
then [Eu/H] should be the same as those of FG stars.
Accordingly, there could be a significant difference in [Eu/H]
even in SG stars.  Therefore,
we use the two terms `polluted' (`p') and `non-polluted' (`np') to discriminate
between stars that are formed from gas polluted by NSM ejecta
and those without such pollution.
For example, $N_{\rm p}$ and $\sigma_{\rm p}$ 
are the number of polluted stars and their [Eu/H] dispersion, respectively.
A basic parameter for the entire distribution of [Eu/H] is
$\sigma_{\rm p}$, $\sigma_{\rm np}$,
$Z_{\rm p}$, $Z_{\rm rp}$,
and $R_{\rm p}$, which is the number ratio of polluted to non-polluted stars:
\begin{equation}
R_{\rm p}= \frac{ N_{\rm p} }{ N_{\rm np} }.
\end{equation}

Figure 5 shows the normalized $N_{\rm p}$ and $N_{\rm np}$ in the four
models with different $R_{\rm p}$, $\sigma_{\rm p}$, and $\sigma_{\rm np}$.
For consistency with observations by W13,
we adopt $Z_{\rm m}$ = $-2.0$ and $-1.6$ for non-polluted
and polluted stars, respectively.
Clearly, the [Eu/H] distributions depend strongly on
the three parameters, with the model with $R_{\rm p}$ being the most
similar to the observed distribution. The relatively large [Eu/H]
dispersion ($0.1$ dex) in the best model with $R_{\rm p}=3.4$ implies
that diffusion of $r$-process elements can proceed efficiently
within the ICM of the GC. The present study cannot discuss whether
such a large dispersion of 0.1 dex can be  achieved in the ICM
though turbulent diffusion. The larger $R_{\rm p}$ implies that 
77\% of the present stars in M15 can be the polluted population. 
The large fraction of the polluted population in M15
implies that the original mass of FG (non-polluted) stars,
from which AGB eject originates,  should be at least by a factor 
of $\sim 10$ larger than the present-day mass of the  FG - This is a classic
mass-budget problem discussed by many previous works (e.g., Smith \& Norris 1982).
The selective stripping of FG stars in the early phase of GC formation
is a promising explanation for solving this mass budget problem (Bekki 2011).

\section{Numerical simulations of gas accumulation from
AGB stars}

In order to discuss whether the density of ICM of forming GCs can 
become as high as $\rho_{\rm icm}=10^4$ cm$^{-3}$, we perform
smooth particle hydrodynamics (SPH)
simulations
of GC formation within massive MCs.
We have already investigated the general trends of GC formation
within MCs (B17b), and we use the same numerical methods used in
B17b in the present study. Since the details of the numerical methods
are given in B17b, we briefly describe them in this paper.
We use our original simulation code that can be run on a cluster
of GPU (Graphics Processing Unit) machines (Bekki 2013, 2015).
A MC with a total mass of $M_{\rm mc}$
and a size of $M_{\rm mc}$
is assumed to have (i) a fractal mass distribution with the three-dimensional
fractal dimension of 2.6 and (ii) a power-law radial density distribution
with a slope of $-1$. The initial virial ratio of a GC-forming MC is
set to be 0.35, which ensures rapid gravitational collapse of the MC.
The initial global rotation of fractal MCs is not included  in the present study.

Feedback effects of SNe with different masses are separately implemented in
the simulations, and the gas ejection of individual AGB stars is also self-consistently
included.
Star formation is assumed to occur (i) if the gas density of a particle ($\rho_{\rm g}$)
exceeds the threshold gas density of 
star formation ($\rho_{\rm th}$)
and (ii) if $div {\bf v}<0$, where $v$ is the velocity of the gas particle.
We adopt 
$\rho_{\rm th}=10^5$ cm$^{-3}$ in the present study
We consider the following two models for star formation from AGB ejecta. 
In one model (M1),  star formation cannot occur from AGB ejecta even 
if $\rho_{\rm g} \ge 10^5 $ cm$^{-3}$. This model is 
constructed to obtain the  better understanding
of the density enhancement achieved through
accumulation of AGB ejecta in the central regions of forming GCs.
In the other model (M2),
 star formation occurs if the above two physical conditions
are met for AGB ejecta.

We here describe the results of the models (M1 and M2) with
$M_{\rm mc}=3 \times 10^6 M_{\odot}$ and $R_{\rm mc}=100$pc for which
GCs with the final stellar mass
($M_{\rm gc}$) being $\sim 10^6 M_{\odot}$ can be formed.
The mass and spatial resolutions of the models are $2.9  M_{\odot}$
and $0.2$pc in the present study.
The total number ($N$) of a simulation significantly increases from 
the initial $N=1048911$
owing to the new addition of `AGB particles' which represent gas ejected
from AGB stars with different masses.

Figures 6, 7, and 8 describe
 how the AGB ejecta can be accreted into the central
region of a forming GC within a giant MC in the fiducial model M1
without secondary star formation from AGB ejecta.
FG stars are formed from numerous small gas clumps
that are  developed from local gravitational instabilities within the fractal MC,
because $\rho_{\rm g}$ of the small clump  can become higher than $10^5$ cm$^{-3}$.
These numerous groups of FG stars merger with one another to form 
a larger stellar system with a stellar halo within a timescale of $10^7$ yr
(at $T=12$ Myr). 
The gas density of the GC-forming MC can dramatically decrease owing
to gas consumption by the FG formation.
The remaining cold gas of the MC can be rapidly expelled
by multiple SN  explosion ($T=12$ and 24 Myr)
because  the lifetimes of massive stars with 
$m > 30 M_{\odot}$ is quite short
($< 10$  Myr).
The cold gas gradually disappears from the inner region of the forming GC
($T=36$ Myr), and almost all of the gas can be expelled from the GC
by $T=110$ Myr.

Massive AGB stars ($m =[6-9] {\rm M}_{\odot}$) start to eject
(Na-rich) gas after the removal of gas chemically polluted by SN explosion
($T \sim 50$ Myr).
The AGB ejecta can be gravitationally trapped in the FG stellar system
with a total mass of $\sim 2 \times 10^6 M_{\odot}$
because of the relatively slow wind velocity ($v_{\rm wind}=10$ km s$^{-1}$).
The gas can be gradually accumulated onto the central regions ($R<2$ pc) of the
FG stellar system ($T=110$ Myr), so that 
$\rho_{\rm g}$ can become very high.
The gas can efficiently lose its kinetic energy 
owing  to energy dissipation of the gas during its accretion process within the GC.
Finally a very compact gaseous sphere in the central region of the FG stellar system
can be formed 
$149$ Myr after the start of gravitational collapse of the MC
in this model. 
These basic formation processes of a GC can be seen in the model M2
with secondary star formation from AGB ejecta.

Figure 9 shows the radial distributions of $\rho_{\rm icm}$ 
at $T=149$ Myr for M1 and M2. Since star formation
is not included in M2, $\rho_{\rm icm}$ can become rather high
$> 10^5$ cm$^{-3}$ in the center of the FG stellar system ($R<0.4$ pc).
Clearly, most of the gas particles within $R<1$ pc has 
$\log \rho_{\rm icm} \ge 3.4$ that is required for $l_{\rm s} \le 3$ pc
(for stopping of $r$-process elements and trapping them within GCs).
This means that if NSMs occur around  this time step ($T=149$ Myr), then 
the ejecta can be easily trapped by the ICM in this model.
It is confirmed that $\rho_{\rm icm}$ can be also quite high 
($\rho_{\rm icm}>10^4$ cm$^{-3}$) at $T=110$ Myr, which means that
fine-tuning in the epoch of an NSM is not required for the ejecta to be trapped
by the ICM.

However, secondary star formation from AGB ejecta can continue to decrease
the total mass of the ICM so that $\rho_{\rm icm}$  can be significantly lower.
Figure 9 shows that although the radial 
gradient of $\rho_{\rm icm}$  is very steep,
$\rho_{\rm icm}$ in the inner region of the 
simulated GC can  still be high ($>10^4$ cm$^{-3}$) 
in M2 with secondary star formation.
This confirms that as long as $\rho_{\rm th}$ is similar to the gas densities
of the molecular cores
of star-forming MCs ($10^5$ cm$^{-3}$), then AGB ejecta (ICM) can 
continue to have $\rho_{\rm icm} \ge 10^4$ cm$^{-3}$ over several
Myrs.
We thus conclude that AGB ejecta can stop the $r$-process elements ejected
from an NSM so that the  NSM ejecta can be used for secondary star formation
in the central region of a forming GC.

\section{Discussion}

\subsection{The epoch of an NSM in a forming M15}

Recent spectroscopic studies of chemical abundances of stars 
have investigated some correlations between light  (Na and O)
and $r$-process elements (Ba and Eu) for M15 (W13 and S11).
Appendix A describes the results of our
own investigation of the correlations using the same data sets.
Clearly, no stars are located in  the area with  [Na/Fe]$>1.0$ and [Eu/H]
$>-1.8$ in M15 whereas there is a strong concentration of  Eu-rich stars 
around [Na/Fe]$\sim 0.3-0.6$. The absence of stars with high [Na/Fe]
and high [Eu/H] might have some physical meaning for the formation of M15.
Since an NSM can occur anytime during GC formation,
it is possible that only  SG stars formed after the NSM can have large
[Eu/H].

Ventura et al. (2011) predicted 
that gas ejected from (i) massive AGB stars with $m=8  M_{\odot}$
(and shorter lifetimes)
can have [Na/Fe]$\sim 1.0$,  whereas those with $m \le 7.5 M_{\odot}$
have lower [Na/Fe] (longer lifetimes).
Therefore,  if an NSM occurs well after gas ejected
from AGB stars  with $m=8 M_{\odot}$ is consumed by secondary star formation,
then the absence of such Na-rich and Eu-rich stars
is expected.
This means that an NSM should have occurred at least $3 \times 10^7$ yr 
(i.e., lifetime of the lowest mass SN) after
the initial burst of star formation in M15.
It is possible that the concentration of [Eu/H] around [Na/Fe]$\sim [0.3-0.5]$
indicates the epoch of an NSM in the early formation phase of M15.
It is observationally unclear whether such an absence of stars with high
[Na/Fe] and high [Eu/H] can be seen in other GCs with internal spreads
of $r$-process elements. Therefore, we cannot make a robust conclusion
on whether the distribution of stars on  the [Na/Fe]-[Eu/H] diagram for a GC
has fossil information on the epoch of a  NSM in the GC.

\subsection{Why do only several GCs show abundance spreads in $r$-process
elements}

R11 investigated the chemical abundances
of $r$-process elements for 17 Galactic GCs, and found that
only 4 GCs among them have large internal spreads of [Eu/H].
So far only these GCs are those with clear evidence of
internal abundance spreads in  $r$-process elements
among all Galactic GCs.
This observation raised a question as to why
only a fraction of GCs show such clear abundance spreads
of [Eu/H] in the Galaxy.
In the present NSM scenario,  NSMs need to occur when forming GCs
still retain an enough amount of AGB ejecta ($>10^3 M_{\odot}$).
Forming GCs can lose their AGB ejecta by some physical processes
such as (i) ram pressure stripping by warm/hot gas of their GC-hosting
galaxies and (ii) prompt  type Ia SNe.
Accordingly, NSMs need to occur before stripping of AGB ejecta
for the ejecta to be converted into new stars.
Given the observed wide range of
the  delay time ($t_{\rm delay}$) distribution of NSMs ($[10^7-10^{10}$] yr),
NSMs do not so easily synchronize with when GCs still retain AGB ejecta.
Therefore, only GCs that happened to experience NSMs during 
secondary star formation from AGB ejecta
can have internal abundance spreads in $r$-process elements.

We  discuss the probability ($R_{\rm r}$)
of one GC to have one NSM in a more quantitative way
as follows. The $t_{\rm delay}$ distribution of NSMs has  the following
 power-law profile
for $t_{\rm detaly} > 3 \times 10^7$ yr (Figure 8 in Dominik et al. 2012):
\begin{equation}
\frac{dN}{dt} \propto t_{\rm delay}^{-\beta},
\end{equation}
where $\beta$ can be approximated as $1.3$. If the duration of secondary
star formation from AGB ejecta is $t_{\rm sf}$, then $P_{\rm r}$ is described
as follows:
\begin{equation}
P_{\rm r}= \frac{ n_{\rm nsm}t_{\rm sf} }{ t_{delay, m} },
\end{equation}
where $n_{\rm nsm}$ is the number of NSMs that can occur in one GC
and $t_{\rm delay, m}$ is the average of $t_{\rm delay}$. 
The total number of GCs with internal abundance spreads in $r$-process elements
in the Galaxy
($N_{\rm gc, r}$) is as follows:
\begin{equation}
N_{\rm gc, r}=  P_{\rm r} N_{\rm gc, 0},
\end{equation}
where $N_{\rm gc,0}$ is the number of the Galactic GCs.
The total number of core collapse SNe (CCSNe) that occur in one GC
(with a total mass of $\approx 10^5 M_{\odot}$) is $\approx 500$,
and one NSM occurs for every 1500 CCSNe (Tsujimoto et al. 2017).
Therefore, $n_{\rm nsm}=500/1500=0.3$ is a reasonable value.
If we adopt reasonable numbers for $t_{\rm delay, m}$ ($=10^9$ yr;
Dominik et al. 2012) and 
$t_{\rm sf}=10^8$ yr and $n_{\rm nsm}=0.3$,  then
$N_{\rm gc, r}=4.5$ for $N_{\rm gc, 0}=150$ (Harris 1996 -- 2010 edition).
This is roughly consistent with the number of GCs with internal
abundance spreads in $r$-process elements.

\subsection{Alternative scenarios}

We have so far assumed that multiple generations of stars formed 
from NSM ejecta (which are diluted by AGB ejecta)
are responsible for the origin of GCs with internal abundance spreads 
in $r$-process elements. 
There are the following two alternative scenarios for the origin.
One is the `gas accretion' scenario in which only  a fraction of
GC stars can accrete  NSM ejecta, so that they can show enhanced
[Eu/H]. The other is a  `merging scenario' in which
two GCs with different [Eu/H] merge with each other to form a new
GC with a [Eu/H] spread.
In the gas accretion scenario, all GC stars were born
almost simultaneously within in a GC-forming MC,
and only some  fraction of  the stars can significantly change
their chemical abundances through accretion of NSM ejecta into the GCs.
Gas accretion occurs for the stars in the core of a GC,
whereas stars outside the core are not affected by gas accretion in this
scenario.
Therefore, there should naturally be the bimodality in the
[Eu/H] distribution, as claimed by TS14.
In order for NSM ejecta to be retained in forming GCs,
a high-density ICM ($\rho_{\rm icm} > 10^6 M_{\odot}$)  is required (TS14).
Although this can be achieved, as shown in M1 of the present
study, a serious problem remains for this scenario.
The observed dispersion of [Na/Fe]
for a given [Eu/H] would be  difficult to explain,
because the gas accretion process does not depend on whether the gas
comprises NSM or AGB ejecta.

In the merger scenario, two GCs with similar masses and [Fe/H] yet
quite different [Eu/H] need to merge with each other.
Even if GC merging is possible in 
dwarf galaxies, as demonstrated by recent numerical
simulations of GC merging (Bekki \& Yong 2012; Bekki \& Tsujimoto 2016),
this scenario has the following problems. 
First, it is unclear why the merging two GCs had almost identical
metallicities yet different [Eu/H] at the time of merging,
if the merging is between two GCs that were 
possibly formed in different parts of the host dwarf galaxy.
Secondly, the merging scenario cannot simply explain why more  Na-rich
stars are more likely to have high [Eu/H] for [Na/Fe]$<1$ in M15.
Thus, the present NSM merger scenario appears to be more promising
than the other two, though it still must resolve the mass budget problem.

\subsection{Relation to the Galactic halo stars}

The Galactic stellar halo contains very metal-poor stars
([Fe/H]$<-2$) with highly enhanced abundances
of $r$-process elements (e.g., Beers \& Chrisllieb 2005).
It remains unclear where
these ``r-I'' ($0.3 \le {\rm [Eu/Fe]} \le 1$ and [Ba/Fe]$<0$)
and ``r-II'' ($1 < {\rm [Eu/Fe]}$ and [Ba/Fe]$<0$) stars originate.
At least some of them  could have been initially
in GCs with large [Eu/H] spreads like M15, and stripped
from the GCs at later times. In addition, it is possible that they
were initially in the building blocks
of the Galaxy, i.e., defunct dwarf galaxies.
Recently, Ji et al. (2016) have discovered the presence of stars with
high [Eu/H] and [Ba/H] in an ultra-faint dwarf galaxy (UFD), Reticulum II
(Ret II), which are identical to r-II stars in the Galactic halo.
Thus, it suggests that r-II stars could also be  from UFDs.
It would be an important question
how we can distinguish between r-II stars from GCs, UFDs, and other types
of defunct dwarf galaxies.

Our NSM scenario predicts that 
some of r-II stars in forming GCs can have high Na abundances.
In fact, most of r-II stars exhibit 
[Na/Fe]$\approx$0.5-1 (Suda et al. 2008), which is compatible with               
their origin associated with GCs like M15.
The above discussion can be used to classify the origin of
individual r-II stars into UFDs or GCs, because r-II stars from UFDs cannot have
such highly enhanced Na abundances. On the other hand, given the lowest
metallicity of $-2.5$ for the Galactic GCs (e.g., Harris 1991),
the r-II stars with [Fe/H]$<-2.5$ in the Galactic halo are unlikely to be
from GCs, whereas the r-II stars in Ret II cover the metallicity range of 
$-3<{\rm [Fe/H]}<-2$ (Ji et al. 2016).
In addition, there are two factors which reduce the contribution to field r-II stars from GCs.
First, more massive GCs are less likely to lose their stars owing to tidal
stripping of the stars during the orbital evolution of GCs around
the Galaxy (e.g., Rossi et al. 2016). Secondly, stars enriched with r-process
elements are predicted to be formed in the central regions of forming GCs. 
Future large samples  of r-II stars in the halo, when combined 
with information on both their metallicity distribution and elemental abundances, 
can reveal the entire picture of r-II stars assembling from GCs, UFDs, or other types of dwarf galaxies.
It should be of noted that carbon-enhanced metal-poor stars with large abundances
of $r$-process elements (CEMP-r) cannot  simply arise from GCs in the present
NSM scenario,
because the SG stars in GCs have highly enhanced N and depleted C, reflecting the 
abundances of massive AGB ejecta (e.g., Bekki et al. 2007).

The origin of r-II stars in Ret II could be related to efficient mixing of interstellar medium (ISM)
and NSM ejecta, and subsequent star formation from the mixed gas
in the early gas-rich phase,  because all stars with [Fe/H]$>-3$ are highly
enriched with r-process elements.
If an NSM is assumed to occur within the ISM with ${\rm n_{HI} }=100$, then
the NSM ejecta can be trapped within the central $\sim 30$ pc of the UFD,  followed by 
star formation from this mixed gas.
If we adopt a shorter $l_{\rm s}$ (stopping length for NSM ejecta)
derived by Tsujimoto et al. (2017), then $l_{\rm s}$ becomes $\sim 4$ pc for
${\rm n_{HI} }=100$. In this case, subsequent star formation is considered to occur after the NSM ejecta 
spread throughout  the ISM due to turbulent diffusion. 
From the observed [Eu/H] spread in Ret II, we can deduce 
the ISM mass ($M{\rm ism}$) mixed with NSM ejecta. Assuming a Eu mass from a single NSM, 
we obtain a large amount of gas of $M{\rm ism} \sim 10^6 M_{\odot}$ (see also the discussion of Ji et al. 2016).
On the other hand, Ret II is known to have the $\gamma$-ray signal likely 
induced by  dark matter annihilation 
(e.g., Geringer-Sameth et al. 2015), 
which leads to the possibility of the high central density of dark matter 
in Ret II. Therefore, it is feasible to consider that the required high ISM density in the central few pc can be realized 
in the early gas-rich Ret II with a high dark matter density.

\section{Conclusions}

In order to understand the origin of GCs with internal abundance spreads
in $r$-process elements (e.g., [Eu/H]),
we have investigated whether ejecta from NSMs can be retained in
forming GCs and subsequently converted into new stars
using analytical models and numerical simulations of GC formation.
In the present `NSM scenario', stars with high [Eu/H] are formed from
NSM ejecta mixed with  ICM (i.e., AGB ejecta) several tens of Myr after
the initial starburst of GC formation. Therefore, prolonged star formation
is essential in this GC formation scenario.
We have also constructed
a theoretical model that explains the observed bimodal distribution of [Eu/H]
in GCs. The principal results are as follows: \\

(1) The high-speed gaseous ejecta (i.e., $r$-process elements)
from NSMs can be stopped by the ICM in forming GCs, if
the total mass of the ICM within the central regions
of GCs ($R \sim [1-3]$ pc) is as large as $10^4 M_{\odot}$.  
This means that AGB ejecta in GCs with the original stellar masses ($M_{\rm s}$)
of $\sim 10^5 M_{\odot}$ can stop the $r$-process elements from NSMs,
because $\sim 10$\% of $M_{\rm s}$  can be AGB ejecta.
It is concluded that AGB ejecta is essentially important for trapping
the NSM ejecta in the central regions of GCs. 
The required larger amount of AGB ejecta implies that
the original GCs should be massive.\\

(2) NSMs can possibly occur more likely outside the 
central region  of a GC than within the central region.
Therefore,  only the fraction of NSM ejecta
can interact with the ICM (i.e., AGB ejecta), so that the mass fraction of the ejecta 
trapped and retained  by the ICM ($f_{\rm ret}$) can be very small ($< 0.1$). 
Our models suggest that small $f_{\rm ret}$ ($\sim 0.01$)  
is required for explaining the observed spreads of [Eu/H] 
(e.g., $\Delta$[Eu/H]$\sim -1$ dex
for M15). Therefore, one NSM in the outer part
of a forming GC is not a problem in the present scenario.  \\

(3) The observed apparent bimodal distribution of [Eu/H] in M15 can be explained
if SG (second generation) stars were formed from ICM with the mean [Eu/H]
being by $\sim 1$ dex higher than the initial [Eu/H] of the GC-forming MC.
The observed large spread in [Eu/H] in the SG stars is due to turbulent diffusion
of NSM ejecta in the ICM. The larger number of SG stars with enhanced [Eu/H]
suggests that the original mass of the GC (M15) should be significantly
larger than the present mass of the GC: this is the classic mass-budget problem
in GC formation. \\

(4) Our new hydrodynamical simulations of GC formation from fractal MCs show 
that the required higher ICM density ($\rho_{\rm icm} > [10^3-10^4]$ cm$^{-3}$)
is possible $\sim 100$ Myr after the initial burst of stars formation in  GCs.
This result thus confirms the validity of the present NSM scenario.
Although secondary star formation can decrease $\rho_{\rm icm}$,
the ICM can keep its higher density owing to slow star formation in SG formation.
We thus conclude that the NSM scenario is very promising in explaining 
the observed internal spreads of $r$-process elements in GCs. \\

(5) The scenario predicts that the number of GCs with such internal spreads
of $r$-process elements should be small ($\le 5$) owing to (i) the rarity of
NSMs and (ii) the very wide distribution of delay time of NSMs.The scenario
also predicts that there are two SG populations with low and high [Eu/Fe],
which appears to be  consistent with the observed [Eu/H]-[Na/Fe] relation for M15.
Although the observed large number of stars with high [Eu/H] and high 
[Na/Fe] in M15 supports
the NSM scenario, other GCs need to be investigated to make a more robust
conclusion on the validity of the scenario. \\

(6) We rule out the possibility of unusual SNe (e.g., magneto-rotational SNe;
Nishimura et al. 2017),
as the origin of internal spreads of $r$-process elements in M15 for the following
two reasons.
First,  if the ICM of forming GCs is chemically polluted by SNe, 
then internal abundance spreads can be seen not only in [Eu/H] but also in
[Fe/H], which is not seen in GCs.
Secondly, 
if SNe are the site of r-process elements,
they produce {\rm both} light (Y, Sr, etc) and heavy (Ba, Eu,
etc) r-process elements.
Stars with enhanced Ba and Eu, however, 
do not exhibit any enhancement in Sr or so in GCs (e.g., M15).
The observed  large scatter seen only
in Ba and Eu (but not in light r-process elements) is inconsistent
with the above (unusual) SN scenario. \\

(7) Since short gamma-ray bursts (GRBs) originate from NSMs
(e.g., Grindlay et al. 2006),
direct evidence for the NSM scenario would be the discovery of short GRBs located
in very young and massive GCs with ages of $[10^7-10^8]$ yr at high redshifts.
Such young GCs can also have very low-level star formation,
and the optical light curves of the GRBs could be significantly
influenced by dust (AGB ejecta) within the GCs.
Although young, massive GCs with short GRBs shrouded by 
dust would be very hard to identify
with current telescopes,
owing to the low luminosities, 
the discovery of such GCs 
would provide irrefutable supporting evidence for the NSM scenario.

\acknowledgments
We are  grateful to the referee, Timothy C. Beers,  for his constructive and
useful comments.
TT is supported in part by JSPS KAKENHI Grant Number 15K05033.

\newpage
\begin{figure}
\epsscale{1.0}
\plotone{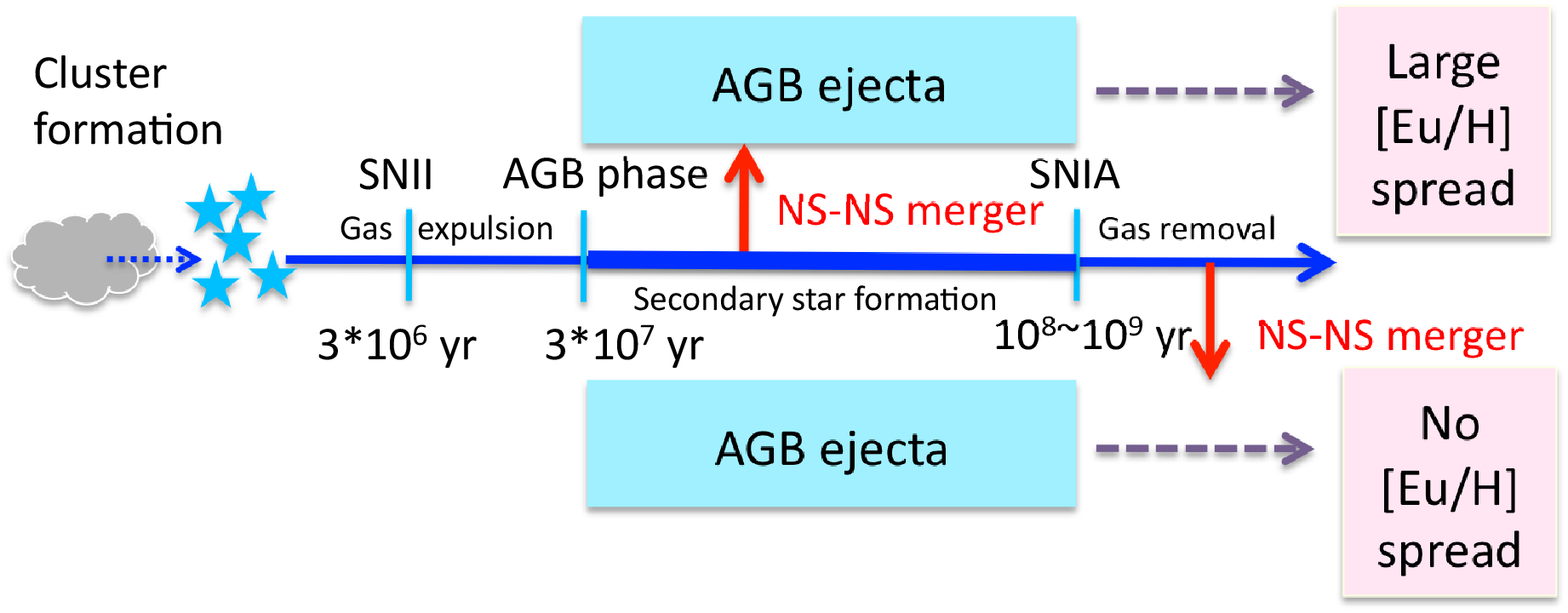}
\figcaption{
A schematic illustration of the new `NSM' scenario of GC formation.
In this scenario,
FG (first generation) stars  are formed from cold gas of a molecular cloud
(MC). After all of the remaining gas left after FG formation is expelled
by Type II SNe of FG ($T=[3-30]\times 10^6$ yr),  gas ejected from AGB stars in FG
stars to be
accumulated into the central region of the forming GC around $T=3 \times 10^7$
yr.  SG (second generation)  stars  can be formed from the gas
when the density of the intra-cluster medium (ICM) becomes as high 
as $10^5$ cm$^{-3}$.  
A merger between neutron stars (`NS-NS' merger; `NSM') occurs
during this SG formation, and subsequently the NSM ejecta can mix with the ICM
(i.e., AGB ejecta).
The mixed gas is converted
into new stars with chemical abundances of $r$-process elements significantly
different from those of FG stars and those of SG stars formed before the NSM.
All of the remaining ICM is removed from the forming GC owing to some physical
processes, such as Type Ia SNe or ram pressure stripping by the GC-host galaxy.
In this scenario,  it depends on the epoch of an NSM in a forming GC
whether the GC can finally have internal abundance spreads in $r$-process elements.
If an NSM occurs in a GC after all of the gas is removed from the GC, then
the GC cannot have such internal abundance spreads.
\label{fig-1}}
\end{figure}
\newpage

\newpage
\begin{figure}
\epsscale{1.0}
\plotone{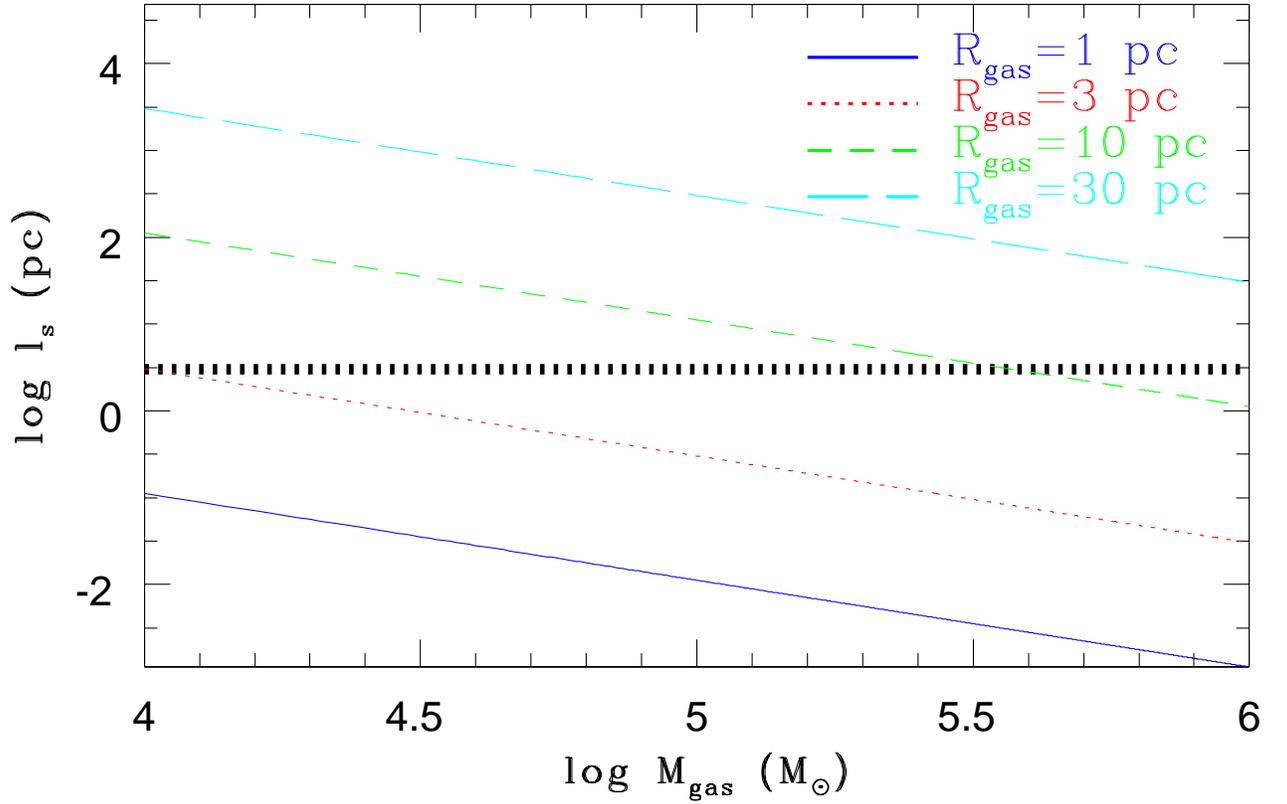}
\figcaption{
Stopping length ($l_{\rm s}$) as a function of the total mass
of gas (AGB ejecta) in a forming GC ($M_{\rm gas}$) 
for $R_{\rm gas}$=1 pc (blue solid), 3 pc (red dotted),
10 pc (green short-dashed), and 30 pc (cyan long-dashed),
where $R_{\rm gas}$ is the radius of the gaseous sphere formed
from the AGB ejecta. If $l_{\rm s}=1$ pc, then it means that a $r$-process
element can be stopped though interaction with gas (mainly
neutral hydrogen) after it freely moves
for 1 pc.
The thick black dotted lines indicates the point where $l_{\rm s}$
becomes the typical radius of GCs ($=3$ pc).
\label{fig-2}}
\end{figure}

\newpage
\begin{figure}
\epsscale{1.0}
\plotone{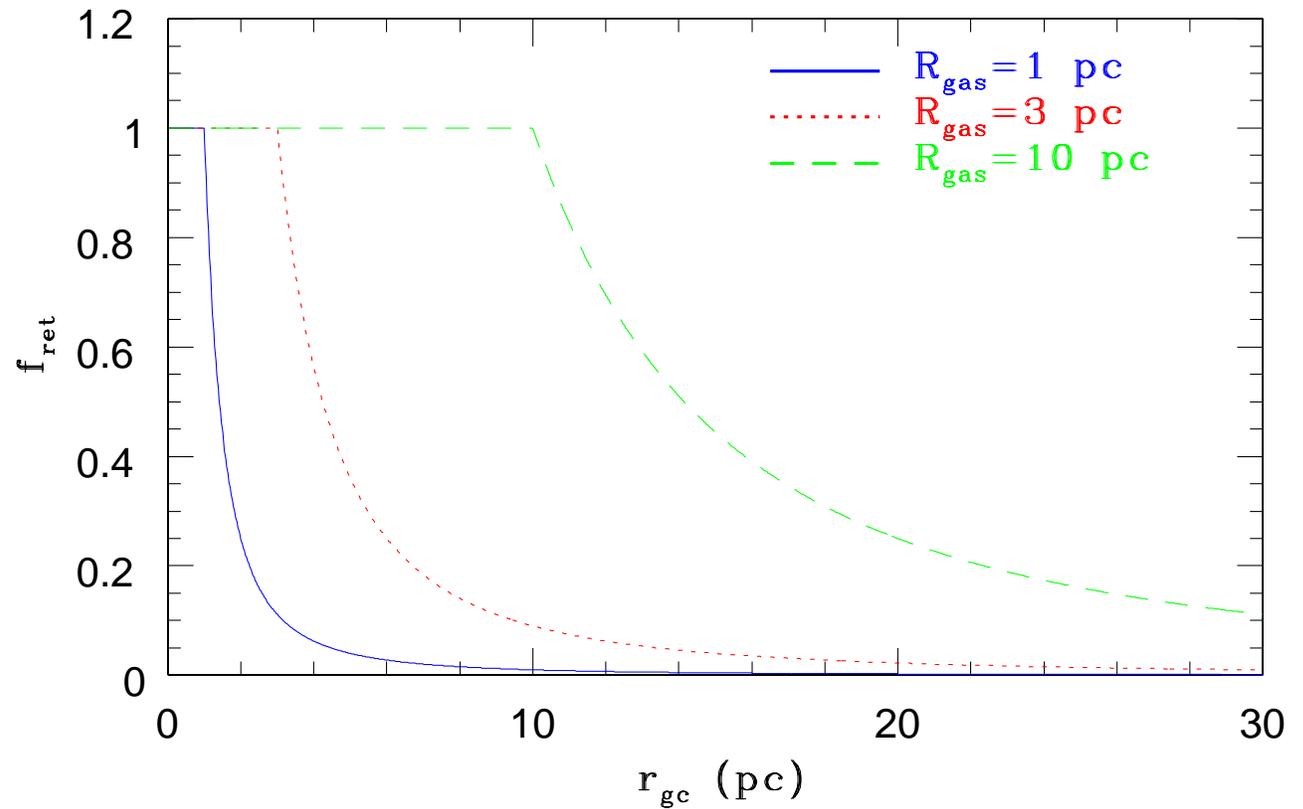}
\figcaption{
Return fraction of NSM ejecta ($f_{\rm ret}$) as a function of
the distance between the location of an NSM and the center of a GC 
($r_{\rm gc}$) for $R_{\rm gas}=1$ pc (blue solid), 3 pc (red dashed),
and 10 pc (green short-dashed).
Since $l_{\rm s}$ is very small ($<10$ pc),
the ejecta
from a  NSM within $R_{\rm gas}$ is assumed to be all trapped by AGB ejecta
in this figure.
\label{fig-3}}
\end{figure}

\newpage
\begin{figure}
\epsscale{1.0}
\plotone{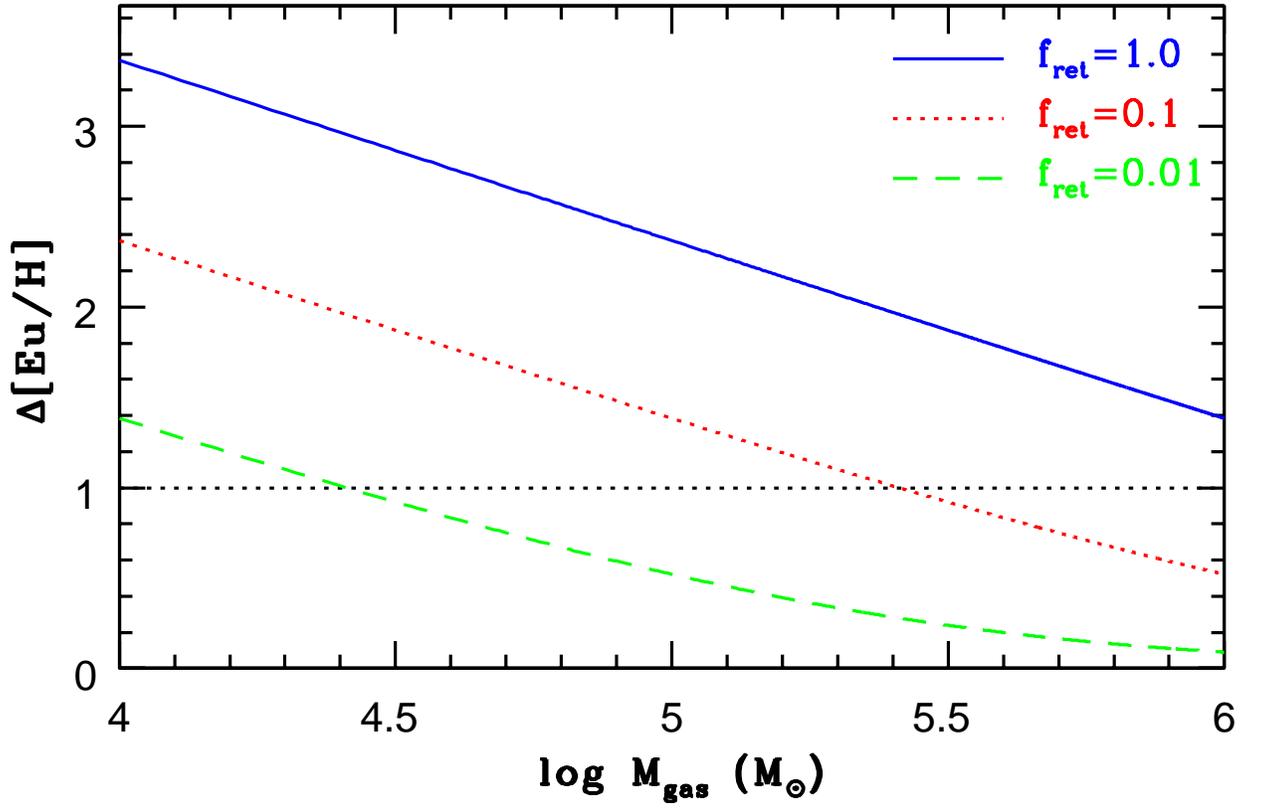}
\figcaption{
Internal spread of [Eu/H] ($\Delta$[Eu/H])
as a function of $M_{\rm gas}$ for $f_{\rm ret}=1.0$ 
(blue solid), 0.1 (red dashed),
and 0.01 (green short-dashed).
The black dotted line indicates the observed ($\Delta$[Eu/H])
for M15 (W13).
\label{fig-4}}
\end{figure}

\newpage
\begin{figure}
\epsscale{1.0}
\plotone{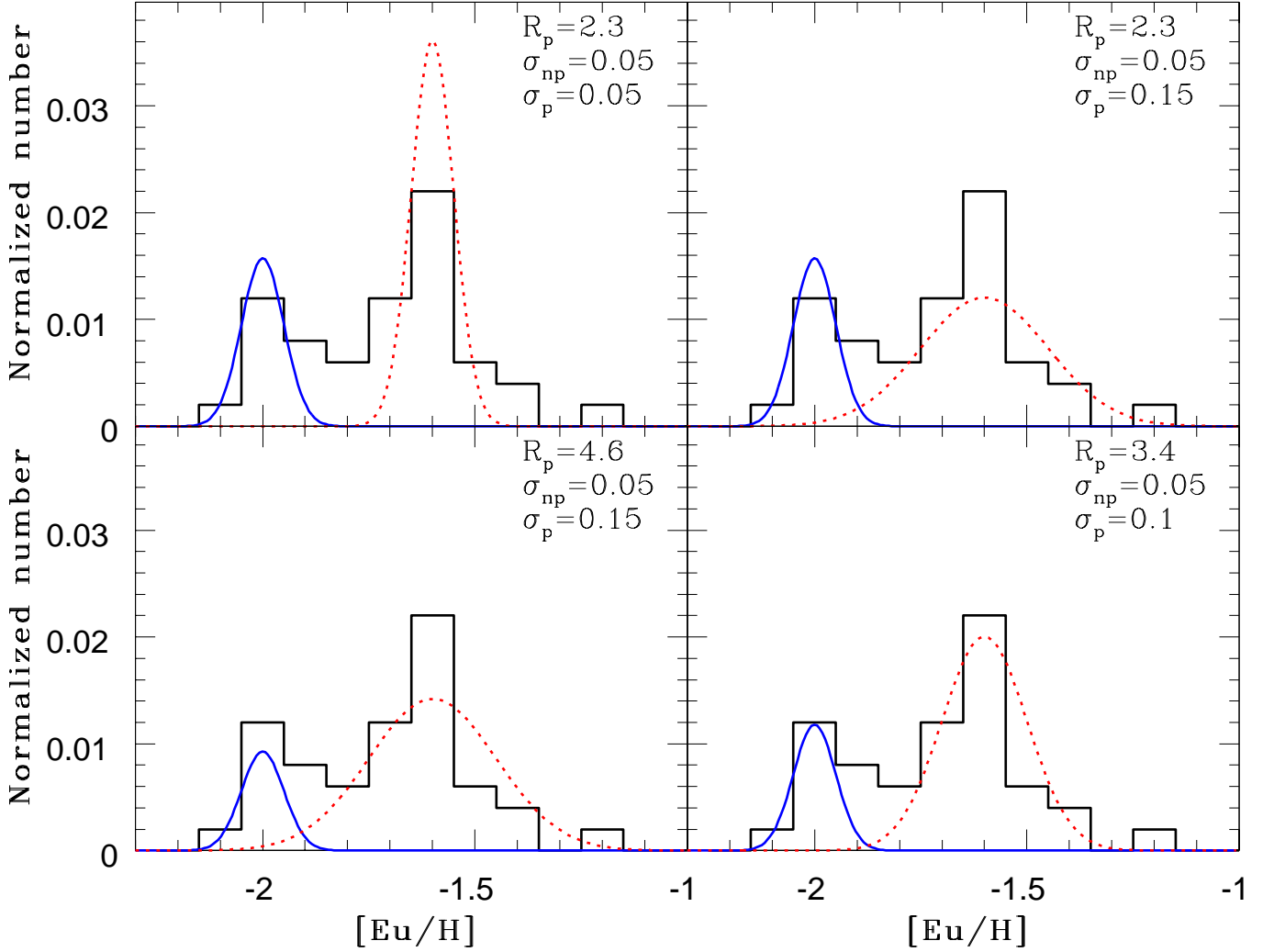}
\figcaption{
Normalized distribution of [Eu/H] for the four representative
models with different $R_{\rm p}$ (number ratio of polluted stars
to non-polluted ones), $\sigma_{\rm np}$ ([Eu/H] dispersion 
for non-polluted stars), and $\sigma_{\rm p}$ ([Eu/H] dispersion
for polluted stars). These values of the parameters are given in
the upper right corner of each panel.
The blue solid and red dotted lines indicate stars formed 
from gas not being polluted by NSM ejecta
and those formed from gas polluted by the ejecta,
respectively.
The thick black line in each panel describes the observed
(normalized) distribution of [Eu/H].
\label{fig-5}}
\end{figure}

\newpage
\begin{figure}
\epsscale{1.0}
\plotone{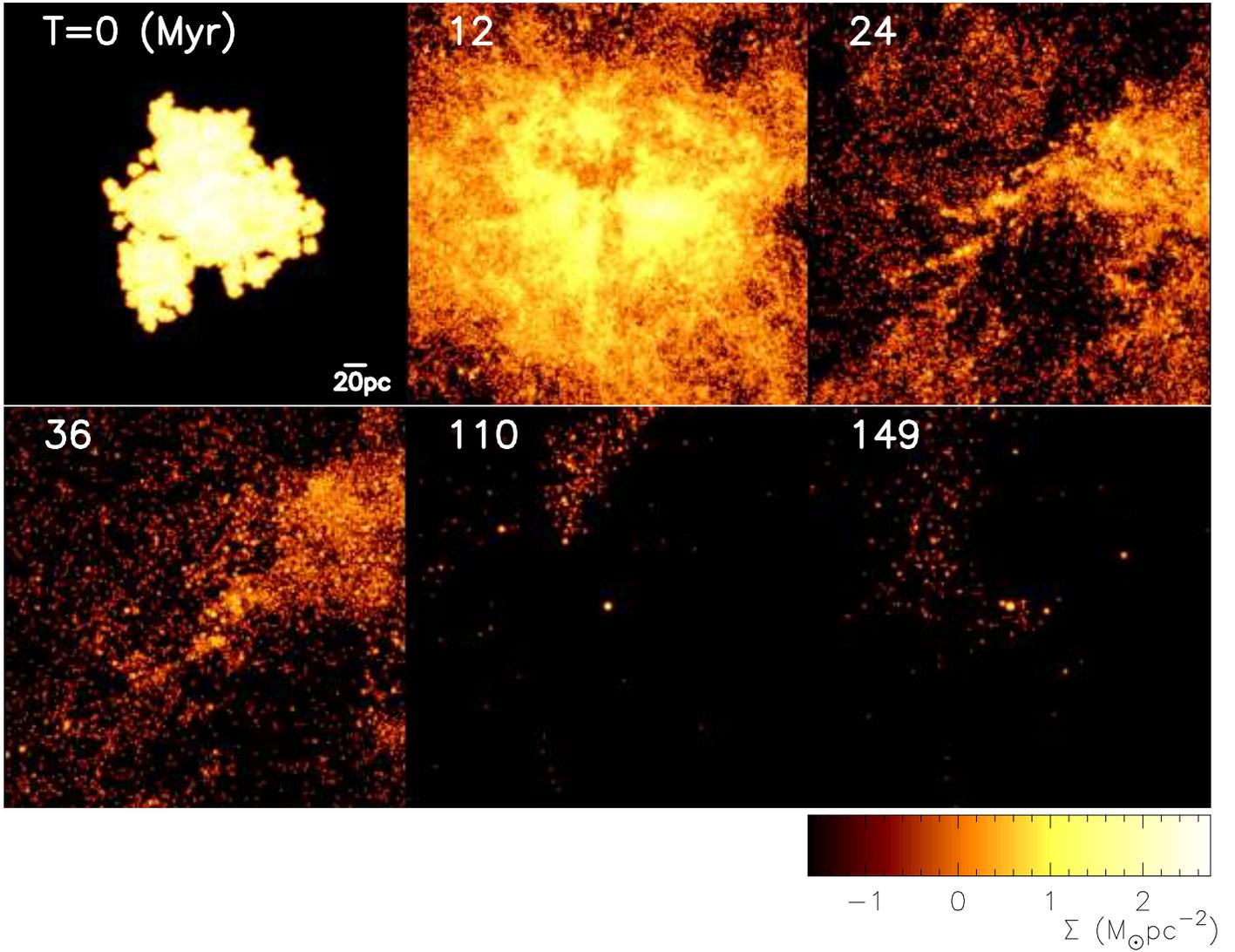}
\figcaption{
Time evolution of the gas density ($\Sigma$)
of  a GC-forming molecular cloud (MC)
projected onto the $x$-$y$ plane for the model M1. Time $T$
(in units of Myr)  is indicated
in the upper left corner of each panel.
The gas density is given in a  logarithmic scale.
\label{fig-6}}
\end{figure}

\newpage
\begin{figure}
\epsscale{1.0}
\plotone{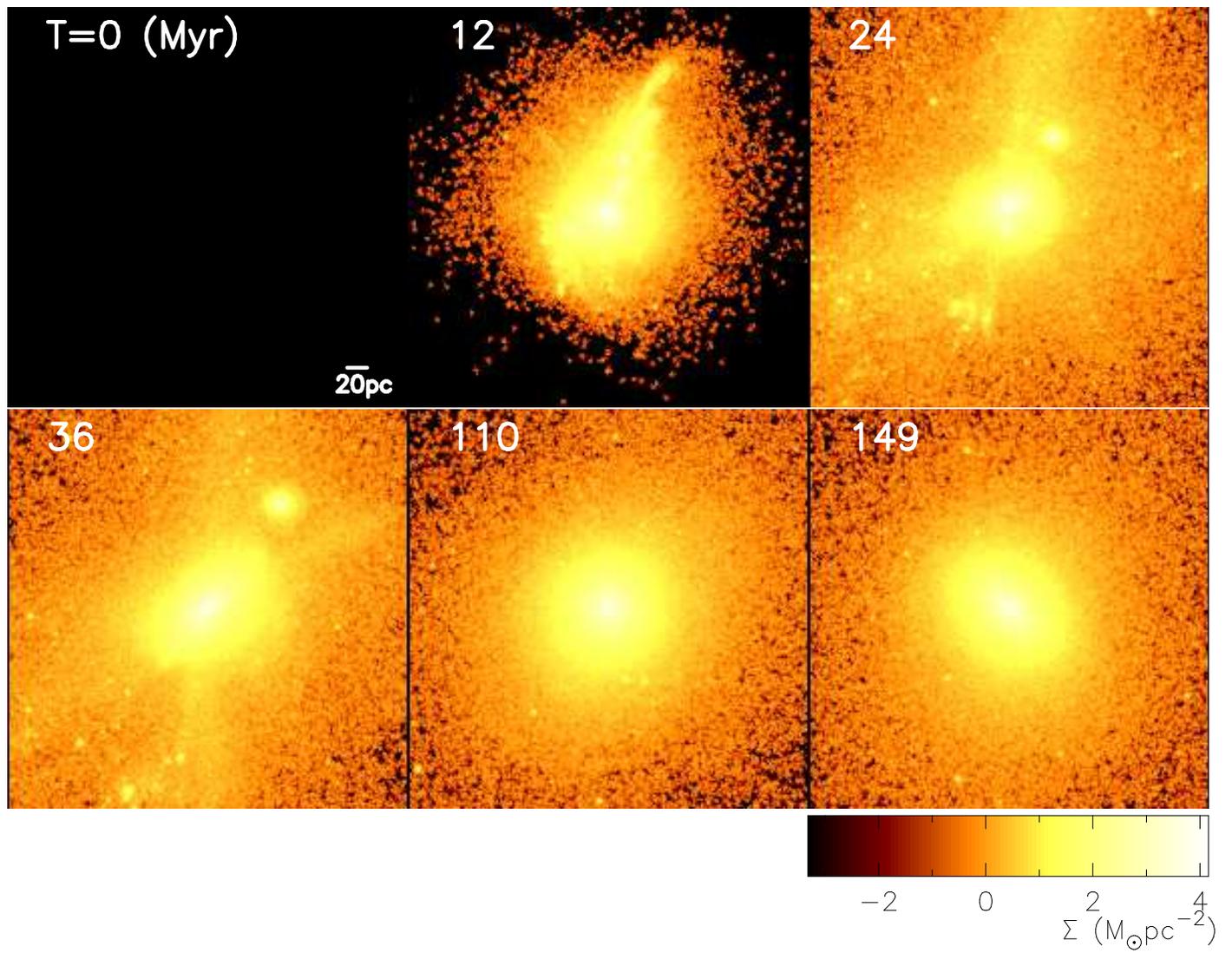}
\figcaption{
The same as Fig. 6 but for the new stars formed from the
initial gas of the MC.
\label{fig-7}}
\end{figure}

\newpage
\begin{figure}
\epsscale{1.0}
\plotone{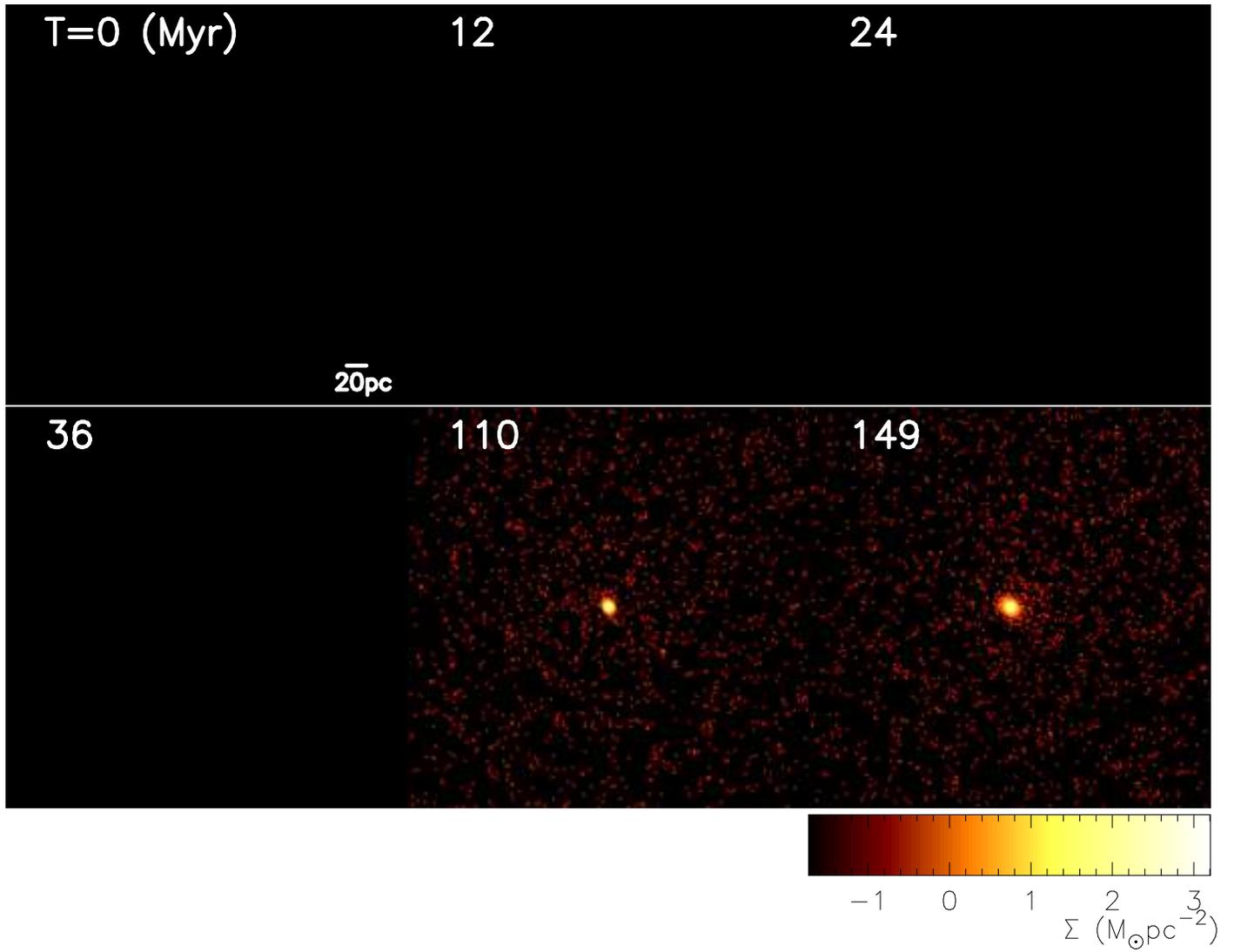}
\figcaption{
The same as Fig. 6 but for the gas ejected from AGB stars.
It is clear that a very high-density compact gaseous region is formed in
the central region 
from AGB ejecta at $T=110$ and 149 Myr.
\label{fig-8}}
\end{figure}

\newpage
\begin{figure}
\epsscale{1.0}
\plotone{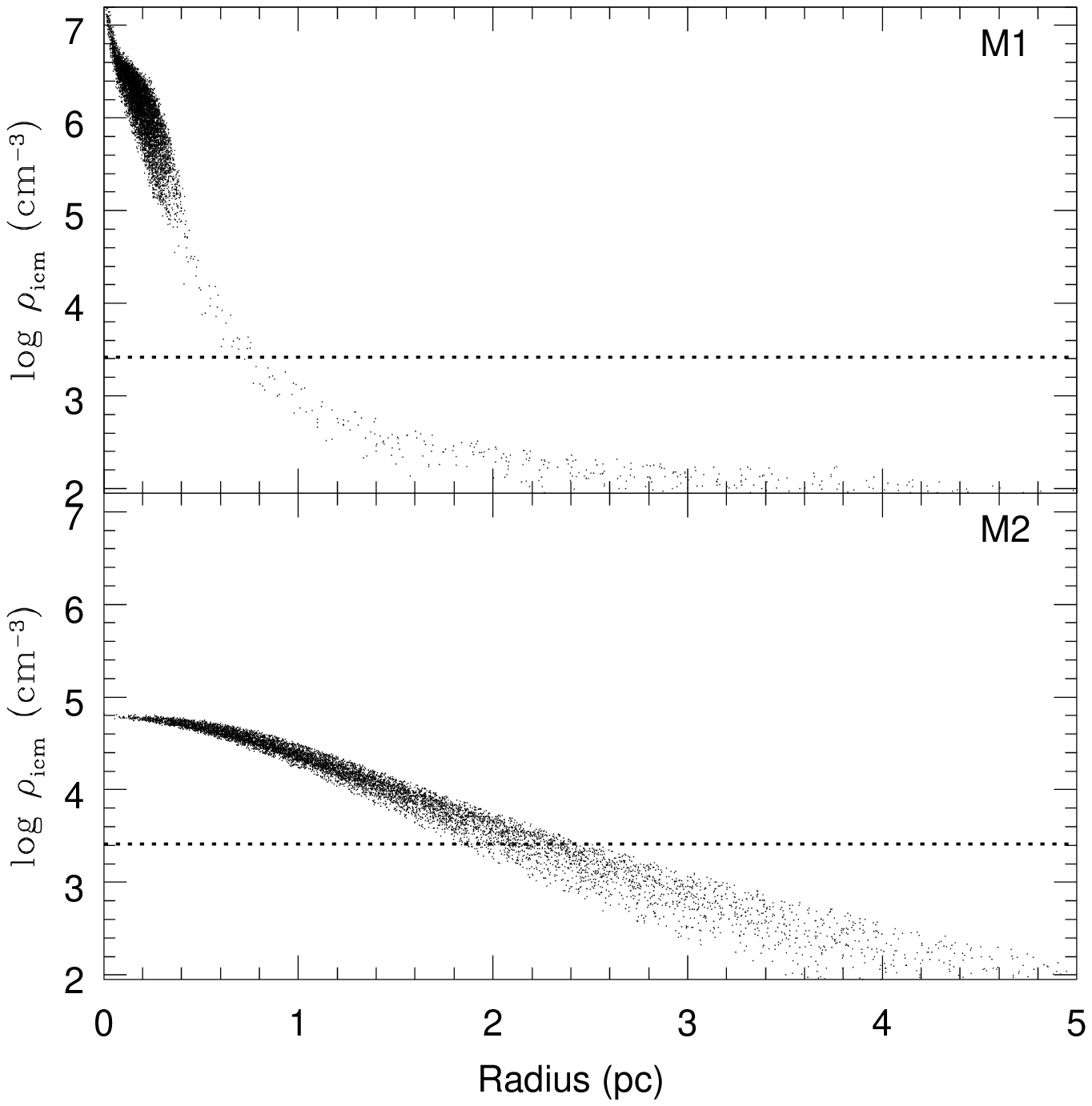}
\figcaption{
Radial distribution of the mass density of the intra-cluster medium 
(ICM; $\rho_{\rm icm}$)
for M1 (upper) and M2 (lower).
Each small dot indicates the location and  $\rho_{\rm icm}$
of each gas particle on this plot.
A horizontal dotted line in each frame
indicates $\rho_{\rm icm}$ above which  NSM ejecta can be stopped
through interaction with the ICM.
M1 and M2 are the GC formation models without  and with secondary
star formation, respectively.
\label{fig-9}}
\end{figure}

\appendix
\section{The observed bimodal distribution of [Eu/H] and correlation
between [Na/Fe] and [Eu/H] in M15}

Recent observational study of the Galactic GC M15 has revealed that
the distributions of [Eu/H] and [Ba/H] ($r$-process elements) 
are bimodal with two distinct peaks (W13). 
We have reproduced the bimodal
[Eu/H] distribution using the data by W13
in order to discuss the present results of our
theoretical models.  Figure A1 shows 
that the [Eu/H] distribution has two peaks around [Eu/H]$\sim -1$ and
$-1.6$, with a large dispersion 
for the entire population ($-2.2 \le {\rm [Eu/H]} \le -1.2$;
$\Delta$[Eu/H]$\sim 1.0$ dex). If one star with very Eu-rich
abundance around [Eu/H]$\sim -1.2$ is removed, then the dispersion is 0.8 dex.
The two peaks strongly suggest that there were two major episodes of star formation
in M15.
The physical origin for the two major episodes of star formation
is given in the main text.

Using the same data for M15, we have investigated the distribution of stars
in the [Na/Fe]-[Eu/H] diagram for M15. Figure A2 shows the following three
trends for the GC stars.  First, most of the stars
with $0.3 \le {\rm  [Na/Fe]} \le 1.0$ (corresponding to SG stars) have
higher [Eu/H] ($>-1.8$, i.e., around the second peak of the [Eu/H]-distribution).  
Secondly, four stars with very high [Na/Fe] ($>1.0$) do not show such high [Eu/H]
($>-1.8$). Thirdly, most of stars with [Na/Fe]$<0$ have lower 
[Eu/H] ($<-1.8$): it should be noted that one
star with [Na/Fe]$\sim -0.1$ has a high [Eu/H] ($\sim -1.5$).
The absence of stars with very high [Na/Fe] and very high [Eu/H] appears to be
remarkable, though the number of stars investigated is not so large.
Gas ejected from metal-poor,  massive AGB stars with $m=8 {\rm M}_{\odot}$ 
can have [Na/Fe]$\sim 1.0$ (e.g., Ventura et al. 2011).
Therefore,
one possible interpretation for the absence of such high-Na and high-Eu stars
is that one NSM occurred after massive AGB stars died away. 
More details on this discussion is given in the main text.

\newpage
\begin{figure}
\epsscale{1.0}
\plotone{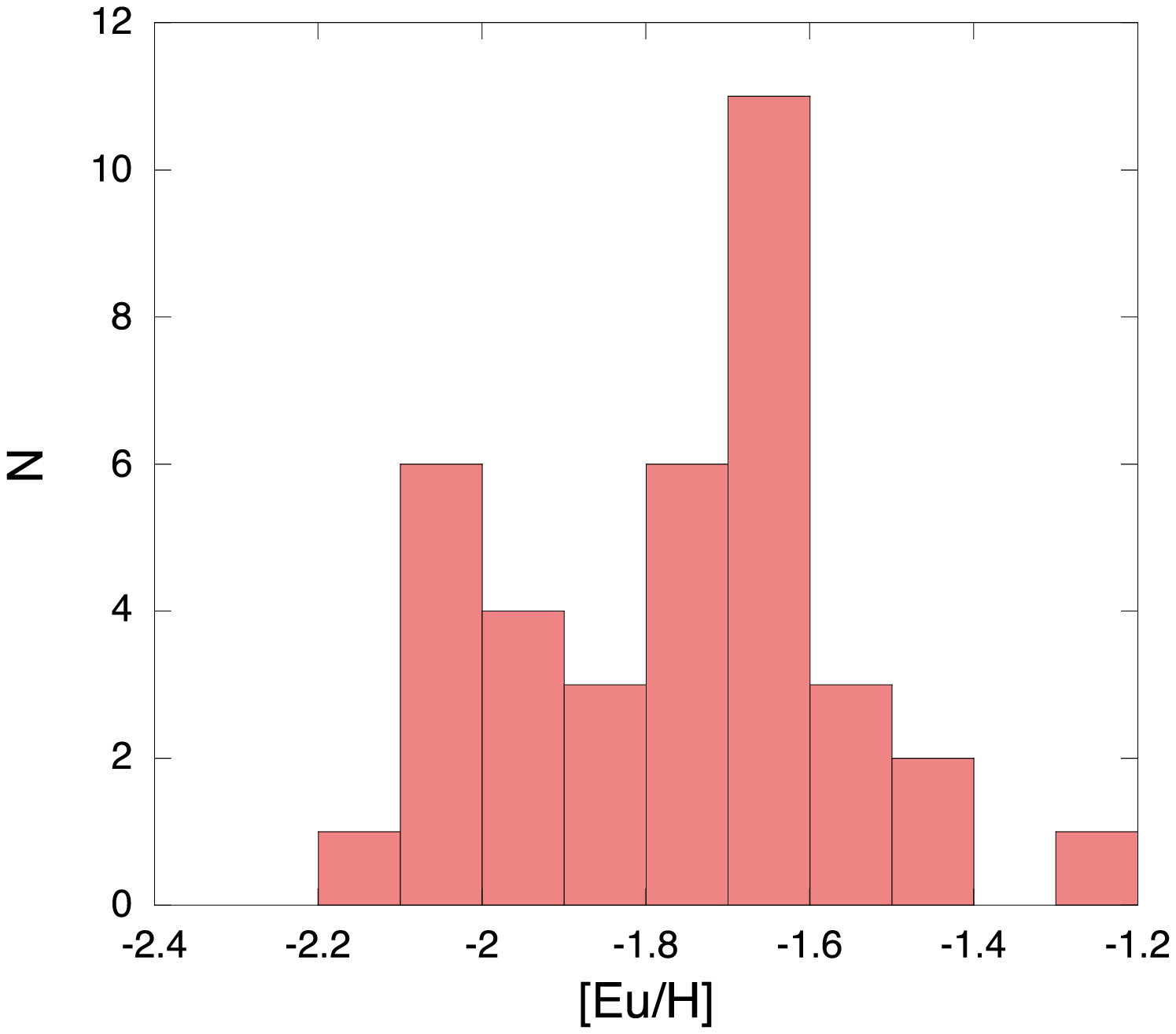}
\figcaption{
[Eu/H] distribution of M15 reproduced from observational data
by W13 and S11 Clearly, there are two peaks around [Eu/H]=$-2.0$
and $-1.6$ in the distribution.
\label{fig-10}}
\end{figure}

\newpage
\begin{figure}
\epsscale{1.0}
\plotone{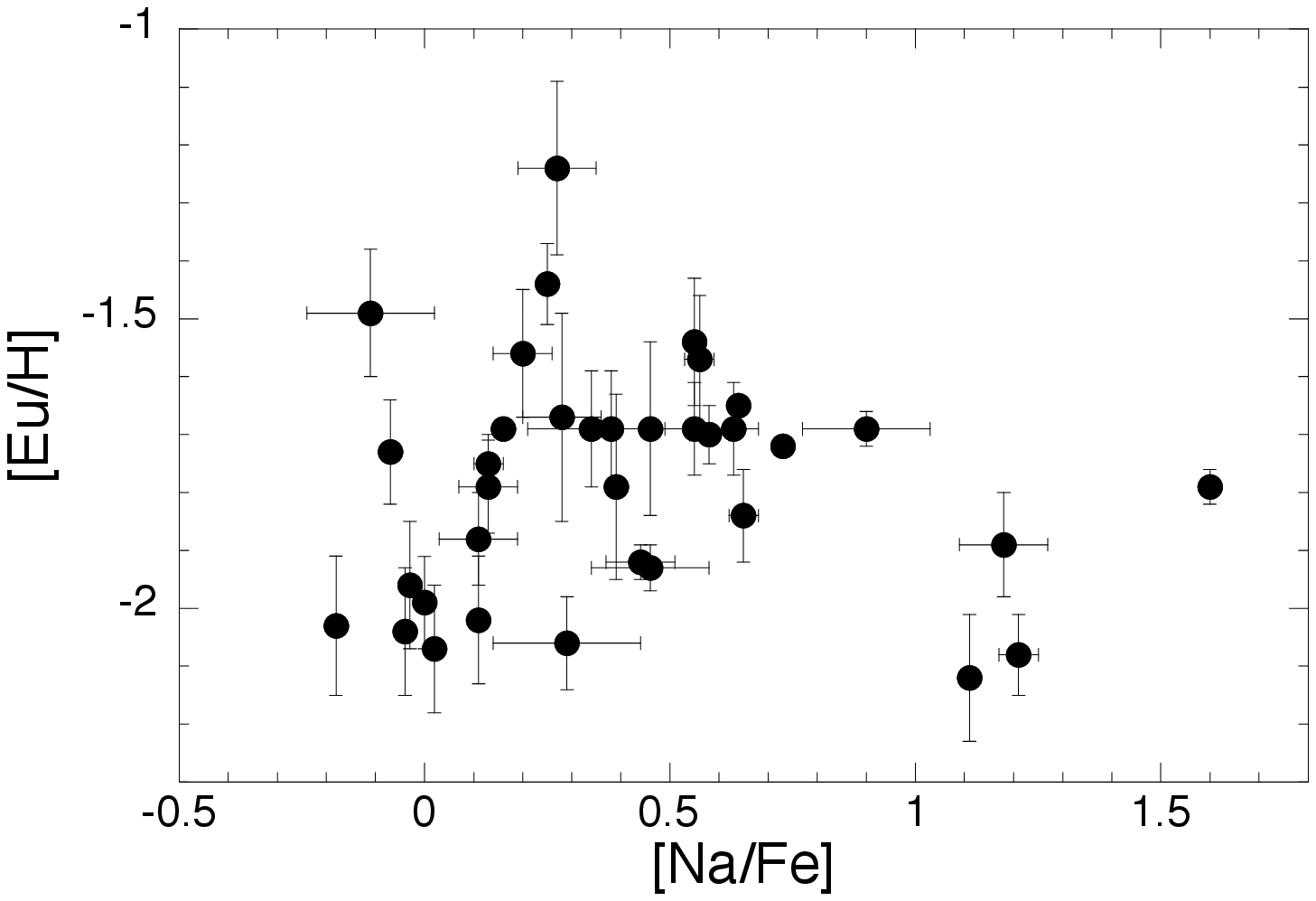}
\figcaption{
[Eu/H] of stars as a function of [Na/Fe] in M15. This plot is reproduced
from the observational data by W13 and S11.
\label{fig-11}}
\end{figure}


\begin{thebibliography}{}

\bibitem[]{}
Beers, T. C., Christlieb, N., 2005, ARA\&A, 43, 531


\bibitem[]{}
Bekki, K., 2011, MNRAS, MNRAS, 412, 2241 

\bibitem[]{}
Bekki, K., 2013, 432, 2298

\bibitem[]{}
Bekki, K. 2015, MNRAS, 449, 1625,

\bibitem[]{}
Bekki, K., 2017a, MNRAS, 467, 1857 (B17a)

\bibitem[]{}
Bekki, K., 2017b, MNRAS in press, arXiv:1705.10039  (B17b)

\bibitem[]{}
Bekki, K., Campbell, S. W., Lattanzio, J. C., Norris, J. E.,
2007, MNRAS, 377, 335

\bibitem[]{}
Bekki, K., Yong, D., 2012, MNRAS, 419, 2063

\bibitem[]{}
Bekki, K., Tsujimoto, T., 2016, ApJ, 831, 70 

\bibitem[]{}
Carretta, E., Bragaglia, A., Gratton, R. G., Lucatello, S.,
2009, A\&A, 505, 117 (C09)

\bibitem[]{}
D'Antona, F., \&  Caloi, V. 2004, ApJ, 611, 871

\bibitem[]{}
D'Ercole, A., Vesperini, E., D'Antona, F., McMillan, S. L. W.,
\& Recchi, S. 2008, MNRAS, 391, 825 (D08)

\bibitem[]{}
D'Ercole, A., D'Antona, F., Ventura, P., Vesperini, E., McMillan, S. L. W., 
2010, MNRAS, 407, 854 (D10)

\bibitem[]{}
Dominik et al. 2012, ApJ, 759, 52

\bibitem[]{}
Fenner, Y., Campbell, S., Karakas, A. I., Lattanzio, J. C.,
Gibson, B. K., 2004, MNRAS, 353, 789

\bibitem[]{}
Freeman, K.,  \& Rodgers, A. W.,  1975, ApJ, 201, 71

\bibitem[]{}
Geringer-Sameth, A., et al. 2015, PhRvL, 115, 1101

\bibitem[]{}
Gratton, Raffaele G.; Carretta, Eugenio; Bragaglia, A.,
2012, A\&ARv, 20, 50

\bibitem[]{}
Greif, T. H., Johnson, J. L., Klessen, R. S., \&  Bromm, V.
2009, MNRAS, 399, 639

\bibitem[]{}
Grindlay, J., Portegies Zwart, S., \&  McMillan, S.
2006, NatPh, 2, 116

\bibitem[]{}
Harris, W. E., 1991, ARA\&A, 29, 543

\bibitem[]{}
Harris, W.E. 1996, AJ, 112, 1487


\bibitem[]{}
Ji, A. P., Frebel, A., Chiti, A., Simon, J. D. 2016, Nature, 531, 610

\bibitem[]{}
Komiya, Y., \& Shigeyama, T. 2016, ApJ, 830, 76

\bibitem[]{}
Larsen, S. S., Brodie, J. P., Grundahl, F., Strader, J.,
2014, ApJ, 797, 15

\bibitem[]{}
Larson, R. B.,
1981, MNRAS, 194, 809

\bibitem[]{}
Marino, A. F., Milone, A. P., Piotto, G., Villanova, S., Bedin, L. R.,
Bellini, A., Renzini, A.,
2009, A\&A, 505, 1099





\bibitem[]{}
Mucciarelli, A.,  Origlia, L.,  Ferraro, F. R.,  Pancino, E.,
2009, ApJ, 695, L134

\bibitem[]{}
Niederhofer, F., et al., 2016, MNARS in press (arXiv:1612.00400)


\bibitem[]{}
Nishimura, N., Sawai, H., Takiwaki, T., Yamada, S., \& Thielemann, F.-K.,
2017, ApJ, 836, L21


\bibitem[]{}
Piotto, G., et al., 2005, 621, 777


\bibitem[]{}
Roederer, I. U., 2011, ApJ, 732, L17 (R11)

\bibitem[]{}
Rossi, L. J., Bekki, K., \& Hurley, J. R., 2016, MNRAS, 462, 2861

\bibitem[]{}
Smith, G. H., \&  Norris, J., 1982, ApJ, 254, 594

\bibitem[]{}
Sneden, C., Kraft, R. P., Shetrone, M. D., Smith, G. H., Langer, G. E., 
\& Prosser, C. F. 1997, AJ, 114, 1964

\bibitem[]{}
Sobeck et al. 2011, AJ, 141, 175 (S11)

\bibitem[]{}
Suda, T., et al. 2008, PASJ, 60, 1159

\bibitem[]{}
Tsujimoto, T., \&  Shigeyama, T. 
2014, ApJ, 795, L18 (TS14)

\bibitem[]{}
Tsujimoto, T., Yokoyama, T., \&  Bekki, K. 2017, ApJ, 835, L3

\bibitem[]{}
Ventura, P., Carini, R., D'Antona, F.,
2011, MNRAS, 415, 3865


\bibitem[]{}
Ventura, P., et al. 2016, ApJ, 831, L17
\bibitem[]{}
Vesperini, E., McMillan, S. L. W.,  D'Antona, F., \& D'Ercole, A.
2010, ApJ, 718, L112

\bibitem[]{}
Worley, C. C., Hill, V., Sobeck, J., \&  Carretta, E.
2013, A\&A, 553, 47 (W13)

\bibitem[]{}
Yong, D. Grundahl, F., \& Norris, J. E.
2015, MNRAS, 446, 3319


\end{thebibliography}
\end{document}